\documentclass[english,preprint]{aastex}
\usepackage[T1]{fontenc}
\usepackage[latin9]{inputenc}
\usepackage{babel}

\begin{document}

\title{The Apparent Velocity and Acceleration of Relativistically Moving Objects}

\author{Austen Berlet$^{1}$, Dennis G. C. McKeon$^{1}$, Farrukh Chishtie$^{1}$,
and Martin Houde$^{2}$}

\affil{$^{1}$Department of Applied Mathematics,}

\affil{$^{2}$Department of Physics and Astronomy,}

\affil{The University of Western Ontario, London, Ontario, Canada N6A 3K7}
\begin{abstract}
Although special relativity limits the actual velocity of a particle
to $c$, the velocity of light, the observed velocity need not be
the same as the actual velocity as the observer is only aware of the
position of a particle at the time in the past when it emits the detected
signal. We consider the apparent speed and acceleration of a particle
in two cases, one when the particle is moving with a constant speed
and the other when it is moving with a constant acceleration. One
curious feature of our results is that in both cases, if the actual
velocity of the particle approaches $c$, then the apparent velocity
approaches infinity when it is moving toward the observer and $c/2$
when it is moving away from the observer.
\end{abstract}

\section{Introduction}

One can derive the relativistic phenomena of the time dilation and length contraction from the Lorentz transformation. However, one does not directly observe these phenomena as the signal used by an observer to detect an event takes a finite length of time to travel from the event to the observer. As a result, two events that occur at the same time in an observer's frame of reference are not actually detected at the same time if they occur at different distances from this observer. This can even result in the apparent velocity of an exceeding $c$, the velocity of light.\\

The consequence of this time delay in receiving the signal of an event has received considerable attention. The appearance of extended objects has been analyzed by a number of authors [Lampe(1924); Penrose(1959); Terrell(1959); Weisskopf(1960); Hickey(1979); Nowojewski(2005); Deissler(2005)]. Astronomical situations in which ``superluminal'' motion has been observed have been discussed in a number of places as well [Rees(1966); Blandford et al.(1977); Blandford and K$\ddot{o}$nigl(1979); Rybicki and Lightman(1979); Mirabel and Rodriguez(1994)]. The possibility of relativistic effects being generated by accelerating astronomical objects has also been considered [Zhou et al.(2004)].\\

In this paper, we discuss pedagogically the apparent velocity and acceleration of an object according to an observer whose observations are affected by the time delay between a signal being emitted by the object and its reception by the observer.

\section{Relativistic Motion at Constant Velocity}

The Lorentz transformation%
\footnote{This transformation was considered before 1900 in studies on the invariance
of the wave equation by \citet{Voigt} and of the Maxwell equations
by \citet{Larmor}.%
}

\begin{equation}
x^{\prime}=\gamma\left(x-vt\right),\,\,\,\, y^{\prime}=y,\,\,\,\, z^{\prime}=z,\,\,\,\, t^{\prime}=\gamma\left(t-\frac{v}{c^{2}}x\right),\label{eq:lorentz}\end{equation}

\noindent where $\gamma=\left(1-v^{2}/c^{2}\right)^{-1/2}$, relates
the space-time coordinates in two coordinate systems $\Sigma$ and
$\Sigma^{\prime}$ moving with a velocity $v$ relative to each other,
provided that $v$ is aligned along the $x$ and $x^{\prime}$ axes
provided they coincide at $t=t^{\prime}=0$. It leads immediately
to such consequences as length contraction, time dilation, and an
upper limit $c$, the speed of light, for the velocity of an object.

However, an observer at time $t$ does not detect the actual position
of a moving particle but only the position of the particle at some
time $\tau$ in the past when it emitted the signal detected at time
$t$. If the signal is a beam of light, then

\begin{equation}
c\left(t-\tau\right)=\left(\mathrm{distance\, traveled\, by\, the\, light\, signal}\right)\label{eq:distance}\end{equation}

\noindent since $c$, the velocity of light, is the same for all observers.
(If the signal were, say, a sound wave, then the velocity of the signal
is dependent on the frame in which it is measured.)

If the particle moving along the $x\mathrm{-axis}$ with speed $v>0$
so that $x=vt$, then a pulse of light emitted at $x_{\mathrm{p}}$
that is received by the observer at $x=0$ satisfies, by equation
(\ref{eq:distance}),

\begin{equation}
c\left(t-\frac{x_{\mathrm{p}}}{v}\right)=\left|x_{\mathrm{p}}\right|.\label{eq:x_p}\end{equation}

\noindent If the particle is receding from the observer, then $x_{\mathrm{p}}>0$
and by equation \ref{eq:x_p}

\begin{equation}
x_{\mathrm{p}}=\frac{vt}{1+\frac{v}{c}};\label{eq:x_p>0}\end{equation}

\noindent while if the particle is approaching the observer, then
$x_{\mathrm{p}}<0$ and

\begin{equation}
x_{\mathrm{p}}=\frac{vt}{1-\frac{v}{c}}.\label{eq:x_p<0}\end{equation}

\noindent As $v\rightarrow c$, the apparent velocity $v_{\mathrm{p}}=dx_{\mathrm{p}}/dt$
approaches $c/2$ for the receding particle and approaches infinity
for the approaching particle. This latter result accounts for some
astronomical objects purportedly having {}``superluminal'' speeds
(see for example \citet{Rees, Rybicki, Deissler}).

In general, the observer is at $\left(d,e,f\right)$ and the particle
at $\left(x_{\mathrm{p}},y_{\mathrm{p}},z_{\mathrm{p}}\right)$ in the rest frame of the observer when
the light signal is emitted detected by the observer at time $t$,
equation (\ref{eq:distance}) then becomes

\begin{equation}
c\left(t-\tau\right)=\left[\left(x_{\mathrm{p}}-d\right)^{2}+\left(y_{\mathrm{p}}-e\right)^{2}+\left(z_{\mathrm{p}}-f\right)^{2}\right]^{1/2}.\label{eq:dist_gen}\end{equation}

If $\left(x_{\mathrm{p}}^{\prime},y_{\mathrm{p}}^{\prime},z_{\mathrm{p}}^{\prime},\tau^{\prime}\right)$ are the coordinates for the emission of the light pulse in the rest frame of the emitter with this frame moving with a velocity $v$ in the $x,x^{\prime}$ direction which coordinates $\left(x_{\mathrm{p}},y_{\mathrm{p}},z_{\mathrm{p}},\tau\right)$ for this event, then,

\noindent From equation (\ref{eq:lorentz})

\begin{equation}
x_{\mathrm{p}}=\gamma\left(x_{\mathrm{p}}^{\prime}+v\tau^{\prime}\right),\,\,\,\, y_{\mathrm{p}}=y_{\mathrm{p}}^{\prime},\,\,\,\, z_{\mathrm{p}}=z_{\mathrm{p}}^{\prime},\,\,\,\,\tau^{\prime}=\gamma\left(\tau-\frac{v}{c^{2}}x_{\mathrm{p}}\right)\label{eq:Lorentz_2}\end{equation}

\noindent so that

\begin{equation}
x_{\mathrm{p}}=\gamma^{-1}x_{\mathrm{p}}^{\prime}+v\tau.\label{eq:xp_gen_a}\end{equation}

\noindent so that equation (\ref{eq:dist_gen}) becomes

\begin{equation}
c\left[t-v^{-1}\left(x_{\mathrm{p}}-\gamma^{-1}x_{\mathrm{p}}^{\prime}\right)\right]=\left[\left(x_{\mathrm{p}}-d\right)^{2}+\left(y_{\mathrm{p}}^{\prime}-e\right)^{2}+\left(z_{\mathrm{p}}^{\prime}-f\right)^{2}\right]^{1/2}.\label{eq:dist_gen_b}\end{equation}

If we solve equation (\ref{eq:dist_gen_b}) for $x_{\mathrm{p}}$, then

\begin{eqnarray}
x_{\mathrm{p}} & = & \gamma^{2}vt+\gamma x_{\mathrm{p}}^{\prime}-\frac{\gamma^{2}v^2}{c^{2}}d\nonumber \\
 &  & \pm\left\{ \left(\frac{\gamma v}{c}\right)^{2}\left(x_{\mathrm{p}}^{\prime}+\gamma vt\right)^{2}+\left(\frac{\gamma^2 v}{c}\right)^{2}\left[\left(\frac{v}{c}d\right)^{2}-2vtd-2\gamma^{-1}dx_{\mathrm{p}}^{\prime}\right]\right.\nonumber \\
 &  & \left.+\left(\frac{\gamma v}{c}\right)^{2}\left[d^{2}+\left(y_{\mathrm{p}}^{\prime}-e\right)^{2}+\left(z_{\mathrm{p}}^{\prime}-f\right)^{2}\right]\right\} ^{1/2}.\label{eq:xp_sol}\end{eqnarray}

If the particle is moving in the direction of increasing $x$ (i.e.,
$v>0$) the negative root is appropriate in equation (\ref{eq:xp_sol}).
In the limit where $x_{\mathrm{p}}^{\prime}=y_{\mathrm{p}}^{\prime}=z_{\mathrm{p}}^{\prime}=d=e=f=0$,
equation (\ref{eq:xp_sol}) reduces to \begin{equation}
x_{\mathrm{p}}=\gamma^{2}v\left(t-\frac{v\left|t\right|}{c}\right),\label{eq:xp_simply}\end{equation}

\noindent which is identical to equations (\ref{eq:x_p>0}) and (\ref{eq:x_p<0}).

Equation (10) has been used \citep{Lampe, Penrose, Terrell, Weisskopf, Scott, Hickey, Nowojewski}
to obtain the apparent shape of an extended object moving with velocity
$v$. From equation (10) it is straightforward to
evaluate the apparent velocity

\begin{eqnarray}
v_{\mathrm{p}} & = & \frac{dx_{\mathrm{p}}}{dt}\nonumber \\
 & = & \gamma^{2}v-\left[\frac{\gamma^{3}v^{3}}{c^{2}}\left(x_{\mathrm{p}}^{\prime}+\gamma vt\right)-\frac{\gamma^{4}v^{3}}{c^{2}}d\right]\nonumber \\
 &  & \cdot\left\{ \left(\frac{\gamma v}{c}\right)^{2}\left(x_{\mathrm{p}}^{\prime}+\gamma vt\right)^{2}+\left(\frac{\gamma^2 v}{c}\right)^{2}\left[\left(\frac{v}{c}d\right)^{2}-2vtd-2\gamma^{-1}dx_{\mathrm{p}}^{\prime}\right]\right.\nonumber \\
 &  & \left.+\left(\frac{\gamma v}{c}\right)^{2}\left[d^{2}+\left(y_{\mathrm{p}}^{\prime}-e\right)^{2}+\left(z_{\mathrm{p}}^{\prime}-f\right)^{2}\right]\right\} ^{-1/2},\label{eq:vp_gen}\end{eqnarray}

\noindent of an object moving with a constant velocity $v$ according
to an observer. From equation (\ref{eq:vp_gen}) it is clear that
this apparent velocity depends explicitly on the position of the observer
$\left(d,e,f\right)$ in its rest frame and the position of the particle
$\left(x_{\mathrm{p}}^{\prime},y_{\mathrm{p}}^{\prime},z_{\mathrm{p}}^{\prime}\right)$
in its rest frame in a non-trivial way.

\section{Relativistic Motion at Constant Acceleration}

We now consider the apparent velocity of a particle that is constantly
accelerated. Non-relativistically, a particle undergoing constant
acceleration eventually reaches an arbitrarily large velocity. Relativistically
an object can never have a velocity exceeding $c$. The very definition
of acceleration is frame dependent on account the Lorentz transformation.
We define {}``constant acceleration'' so that all observers in inertial
reference frames who instantaneously see the particle at rest (i.e.,
{}``comoving'' with the particle) see the same acceleration.

If the observer is in a frame $\Sigma$ in which the particle is moving
along the $x$-axis, then it is easy to verify that an equation that
satisfies the above criterion for acceleration is

\begin{equation}
x^{2}-c^{2}t^{2}=\alpha^{2},\label{eq:hyperbola}\end{equation}

\noindent where $\alpha$ is a constant. It follows from equation
(\ref{eq:hyperbola}) that the particle moves along one of two branches
of an hyperbola in the $x\mathrm{-}t$ plane; we take the particle
to be moving along the branch on which $x>0$. If $t<0$, the particle
is approaching the origin (at $x=0$), while for $t>0$ it is receding
from it. Any light signal emitted form the particle can only be detected
at $x=0$ at a time $t>0$, as $x=0$ is causally disconnected from
any event on the trajectory of equation (\ref{eq:hyperbola}) where
$t\leq0$.

To verify that equation (\ref{eq:hyperbola}) satisfies the criterion
for constant acceleration, we note that it implies that

\begin{equation}
\frac{dx}{dt}=\frac{c^{2}t}{\sqrt{\alpha^{2}+c^{2}t^{2}}}\label{eq:dxdt}\end{equation}

\noindent and

\begin{equation}
\frac{d^{2}x}{dt^{2}}=\frac{c^{2}}{\sqrt{\alpha^{2}+c^{2}t^{2}}}-\frac{c^{4}t^{2}}{\left(\alpha^{2}+c^{2}t^{2}\right)^{3/2}}.\label{eq:d2xdt2}\end{equation}

By equation (\ref{eq:dxdt}), we see that the instant in $\Sigma$
when the particle is at rest is $t=0$; from equation (\ref{eq:d2xdt2})
we see that at that instant the acceleration is $c^{2}/\alpha$. We
note that if $\lambda$ is the {}``proper time'' (i.e., $d\lambda=\sqrt{1-v^{2}/c^{2}}dt$),
then $dx/d\lambda=tc^{2}/\alpha$.

Now if $\Sigma^{\prime}$ is an inertial observer whose coordinates
are related to those of $\Sigma$ by a Lorentz transformation (as
defined in eq. (1)), it follows that

\[
x^{2}-c^{2}t^{2}=x^{\prime2}-c^{\prime2}t^{\prime2},\]

\noindent so that the trajectory of the particle as viewed by an observer
in $\Sigma^{\prime}$ is

\begin{equation}
x^{\prime2}-c^{\prime2}t^{\prime2}=\alpha^{2}.\label{eq:hyperbolaprime}\end{equation}

Since equations (\ref{eq:hyperbola}) and (\ref{eq:hyperbolaprime})
are of the same form, it follows immediately that in $\Sigma^{\prime}$,
the particle is also at rest at $t^{\prime}=0$ and that at that instant
its acceleration is $c^{2}/\alpha$. Consequently all comoving observers
would see the particle as moving with this acceleration. It is apparent
from equation (\ref{eq:dxdt}) that the particle always moves with
a velocity less than $c$ provided that $\alpha^{2}>0$, and that
as $t\rightarrow\pm\infty$ this velocity approaches $\pm c$.

Let us now consider an observer located at $x=0$ receiving a light
signal at time $t$ emitted by a particle moving along the trajectory
given by equation (\ref{eq:hyperbola}) at time $\tau$. If this particle
is at $x_{\mathrm{p}}$ at $\tau$, then together equations (\ref{eq:distance})
and (\ref{eq:hyperbola}) imply that

\begin{equation}
c\left[t-\left(\pm\frac{\sqrt{x_{\mathrm{p}}^{2}-\alpha^{2}}}{c}\right)\right]=\left|x_{\mathrm{p}}\right|,\label{eq:|xp|}\end{equation}

\noindent where the positive (negative) root occurs if the particle
is receding from (approaching) the observer. As stated before, we
consider only the case in which the particle is moving along the branch
of the hyperbola for which $x_{\mathrm{p}}>0$. With this in mind,
it follows from equation (\ref{eq:|xp|}) that

\begin{equation}
x_{\mathrm{p}}=\frac{ct}{2}+\frac{\alpha^{2}}{2ct},\label{eq:xp_a}\end{equation}

\noindent which also implies that $0<t<\infty$. The apparent velocity
of the particle is given by

\begin{eqnarray}
v_{\mathrm{p}} & = & \frac{dx_{\mathrm{p}}}{dt}\nonumber \\
 & = & \frac{c}{2}-\frac{\alpha^{2}}{2ct^{2}}.\label{eq:vp_c}\end{eqnarray}

\noindent We, therefore, see that as $t\rightarrow\infty$, $v_{\mathrm{p}}\rightarrow c/2$
and as $t\rightarrow0^{+}$, $v_{\mathrm{p}}\rightarrow-\infty$.
This is consistent with results following from equations (\ref{eq:x_p>0})
and (\ref{eq:x_p<0}).

We now return to the situation in which the particle moves along the
trajectory of equation (\ref{eq:hyperbola}), but we now position
our observer at $\left(\xi,0,\zeta\right)$ in his rest frame, rather than at $(0,0,0)$. If again, the particle
emits a light signal at time $\tau$ when it is at position $\left(x_{\mathrm{p}},0,0\right)$,
we find that from equation (\ref{eq:distance})

\begin{equation}
c\left(t-\tau\right)=\left[\left(x_{\mathrm{p}}-\xi\right)^{2}+\zeta^{2}\right]^{1/2},\label{eq:dist_gen_c}\end{equation}

\noindent where $t$ is the time the signal arrives at $\left(\xi,0,\zeta\right)$.
From equation (\ref{eq:hyperbola}) we find that

\begin{equation}
\tau=\pm\frac{\sqrt{x_{\mathrm{p}}^{2}-\alpha^{2}}}{c},\label{eq:tau}\end{equation}

\noindent so that equation (\ref{eq:dist_gen_c}) becomes

\begin{equation}
t=c^{-1}\left\{ \pm\left(x_{\mathrm{p}}^{2}-\alpha^{2}\right)^{1/2}+\left[\left(x_{\mathrm{p}}-\xi\right)^{2}+\zeta^{2}\right]^{1/2}\right\} .\label{eq:t}\end{equation}

As a result,

\begin{eqnarray}
v_{\mathrm{p}} & = & \frac{dx_{\mathrm{p}}}{dt}\nonumber \\
 & = & c\left\{ \pm\frac{x_{\mathrm{p}}}{\left(x_{\mathrm{p}}^{2}-\alpha^{2}\right)^{1/2}}+\frac{\left(x_{\mathrm{p}}-\xi\right)}{\left[\left(x_{\mathrm{p}}-\xi\right)^{2}+\zeta^{2}\right]^{1/2}}\right\} ^{-1},\label{eq:vp_final}\end{eqnarray}

\noindent which is just the $\xi=\zeta=0$ limit of equation (19),
as expected. Equation (\ref{eq:xp_final}) reduces to equation (\ref{eq:xp_a}) once $\xi=\zeta=0$.
To express $x_{\mathrm{p}}$ in terms of $t$, one inverts
equation (\ref{eq:t}) to yield

\begin{equation}
x_{\mathrm{p}}=\frac{\xi\left(\xi^{2}+\zeta^{2}+\alpha^{2}-c^{2}t^{2}\right)-f(t)}{2\left(\xi^{2}-c^{2}t^{2}\right)}\label{eq:xp_final}\end{equation}

where
\begin{equation}
f(t)=\sqrt{4\left(c^{2}t^{2}-\xi^{2}\right)c^{2}t^{2}\alpha^2
+c^2t^2\left(\xi^{2}+\zeta^{2}+\alpha^{2}-c^{2}t^{2}\right)^{2}}
\end{equation}

Hence,
\begin{eqnarray}
v_{\mathrm{p}} & = & \frac{dx_{\mathrm{p}}}{dt}\nonumber \\
 & = & \frac{-\left(\xi^2-c^2t^2\right)f'(t)+2c^2t\left(\xi\zeta^2+\xi\alpha^2-f(t)\right)}
 {2\left(\xi^{2}-c^{2}t^{2}\right)^2}
\end{eqnarray}

and the apparent acceleration can also be computed as follows,
\begin{eqnarray}
a_{\mathrm{p}} & = & \frac{dv_{\mathrm{p}}}{dt}\nonumber
\end{eqnarray}

\begin{eqnarray}
=
\frac{-\left(\xi^2-c^2t^2\right)^2f''(t)+\left[2c^2(\xi^2-c^2t^2)+6c^4t^2\right]\left[\xi\zeta^2+\xi\alpha^2-f(t)\right]
-4c^2tf'(t)\left(\xi^2-c^2t^2\right)}{{2\left(\xi^{2}-c^{2}t^{2}\right)^3}}
\end{eqnarray}

where

\begin{equation}
f'(t)=\frac{c^2t}{f(t)}\left[(\xi^2+\zeta^2+\alpha^2-c^2t^2)(\xi^2+\zeta^2+\alpha^2-3c^2t^2)-4\alpha^2\xi^2+8c^2t^2\alpha^2\right]
\end{equation}

and $f''(t)$ can be computed in a similar manner.

From eq. (24) it follows that we have the following asymptotic limits

(1) As $t\rightarrow\infty$ ($\xi$, $\zeta$ fixed),

\begin{equation}
v_p\rightarrow\frac{c}{2}+\frac{\zeta^2-\alpha^2}{2ct^2}+\frac{\xi(\alpha^2+\zeta^2)}{c^2t^3}+O\left(\frac{1}{t^4}\right)
\end{equation}

\begin{equation}
a_p\rightarrow\frac{\alpha^2-\zeta^2}{2ct^2}-\frac{3\xi(\alpha^2+\zeta^2)}{c^2t^4}+O\left(\frac{1}{t^5}\right)
\end{equation}

(2) As $\xi\rightarrow\infty$ ($t$, $\zeta$ fixed),

\begin{equation}
v_p\rightarrow-\frac{c}{2}+\frac{c(\alpha^2-\zeta^2)}{2\xi^2}+\frac{c^2t(\alpha^2+\zeta^2)}{\xi^3}+\frac{3c^3t^2(\alpha^2-\zeta^2)-2\zeta^2\alpha^2c}{2\xi^4}+\frac{2c^4t^3(\zeta^2+\alpha^2)}{\xi^5}+O\left(\frac{1}{\xi^6}\right)
\end{equation}

\begin{equation}
a_p\rightarrow\frac{c^2(\alpha^2+\zeta^2)}{\xi^3}+\frac{3c^3t(\alpha^2-\zeta^2)}{\xi^4}+O\left(\frac{1}{\xi^5}\right)
\end{equation}

(3) As $\zeta\rightarrow\infty$ ($t$, $\xi$ fixed),

\begin{equation}
v_p\rightarrow-\frac{c}{2}\left[\frac{\zeta^2}{(\xi+ct)^2}+\left(1+\frac{\alpha^2}{(\xi+ct)^2}\right)\right]+O\left(\frac{1}{\zeta^2}\right)
\end{equation}

\begin{equation}
a_p\rightarrow\frac{c^2(\alpha^2+\zeta^2)}{(\xi+ct)^3}+O\left(\frac{1}{\zeta^2}\right)
\end{equation}

We see from eqs.(29-34) that the limits $t\rightarrow\infty$, $\xi\rightarrow\infty$ and $\zeta\rightarrow\infty$ do not commute. Furthermore, as $\zeta\rightarrow\infty$ both $v_p$ and $a_p$ diverge. It would be of interest to consider astronomical situations that would serve to illustrate these results.

\acknowledgements{G. McKeon would like to thank F. Brandt for helpful discussions and
R. Macleod for a suggestion.}

\affil{}
\end{document}